\documentclass[12pt]{article}

\usepackage{amsfonts,amsmath,amssymb,accents,mathrsfs}
\usepackage[retainorgcmds]{IEEEtrantools}
\usepackage{multirow}
\usepackage{multicol}
\usepackage{tikz}
\usepackage{graphicx,color}
\usepackage{booktabs}
\usepackage{placeins}
\usepackage[colorlinks,linkcolor=Blue,citecolor=Blue,urlcolor=Blue,bookmarks,bookmarksnumbered]{hyperref}
\usepackage[nosort]{cite}
\usepackage[scaled=0.85]{helvet}
\usepackage[utf8]{inputenc}
\usepackage{cancel}

\definecolor{Red}    {rgb}{0.90,0.00,0.12} %  1
\definecolor{Blue}   {rgb}{0.00,0.00,1.00} %  2
\definecolor{Green}  {rgb}{0.10,0.70,0.10} %  3
\definecolor{Turque} {rgb}{0.00,0.65,0.85} %  4
\definecolor{Orange} {rgb}{1.00,0.50,0.15} %  5
\definecolor{Magenta}{rgb}{1.00,0.00,1.00} %  6
\definecolor{Gold}   {rgb}{1.00,0.75,0.25} %  7
\definecolor{Seaweed}{rgb}{0.01,0.24,0.09} %  8
\definecolor{Purple} {rgb}{0.50,0.25,0.55} %  9
\definecolor{Brown}  {rgb}{0.43,0.26,0.32} % 10
\definecolor{grey1}  {rgb}{0.20,0.20,0.20} % 11
\definecolor{grey2}  {rgb}{0.40,0.40,0.40} % 12
\definecolor{grey3}  {rgb}{0.60,0.60,0.60} % 13
\definecolor{grey4}  {rgb}{0.80,0.80,0.80} % 14
\definecolor{grey5}  {rgb}{0.90,0.90,0.90} % 15

\def\a{{\alpha}}
\def\b{{\beta}}
\def\g{{\gamma}}
\def\d{{\delta}}

\def\th{{\theta}}

\def\ad{{\dot{\alpha}}}
\def\bd{{\dot{\beta}}}
\def\gd{{\dot{\gamma}}}

\def\thd{{\bar{\theta}}}

\def\N{{\mathcal{N}}}

\def\J{{\mathcal{J}}}
\def\T{{\mathcal{T}}}

\def\W{{\mathcal{W}}}

\def\D{{\rm D}}
\def\Dd{{\bar{\rm D}}}
\def\pa{\partial}

\def\be{\begin{equation}}
\def\ee{\end{equation}}
\def\bea{\begin{IEEEeqnarray*}}
\def\eea{\end{IEEEeqnarray*}}
\def\n{\IEEEyesnumber}
\def\sn{\IEEEyessubnumber}

\makeatletter
\def\section{\@startsection{section}{1}{\z@}
              {3ex plus-1ex minus-.2ex}{1pt plus1pt}{\large\sf\bfseries\boldmath}}
\def\subsection{\@startsection{subsection}{2}{\z@}
              {1.5ex plus-1ex minus-.2ex}{0.01pt plus1pt}{\sf\slshape}}
\def\subsubsection{\@startsection{subsubsection}{3}{\z@}
              {1.5ex plus-1ex minus-.2ex}{0.01pt plus0.2pt}{\sf\boldmath}}
\def\paragraph{\@startsection{paragraph}{4}{\z@}
              {.75ex \@plus.5ex \@minus.2ex}{-2mm}{\sf\bfseries\boldmath}}
\makeatother

   \parskip\medskipamount     \lineskip=0pt
\thicklines
\begin{document}
\thispagestyle{empty}
\noindent{\small
\hfill{HET-1766 {~} \\ % un-comment-out and specify when done}
$~~~~~~~~~~~~~~~~~~~~~~~~~~~~~~~~~~~~~~~~~~~~~~~~~~~~~~~~~~~~$
$~~~~~~~~~~~~~~~~~~~~\,~~~~~~~~~~~~~~~~~~~~~~~~~\,~~~~~~~~~~~~~~~~$
 {~}
}
\vspace*{6mm}
\begin{center}
{\large \bf 
Conserved higher spin supercurrents for arbitrary spin massless supermultiplets
and higher spin superfield cubic interactions
\vspace{3ex}
} \\   [9mm] {\large { I. L.
Buchbinder,\footnote{joseph@tspu.edu.ru}$^{a,b,c}$ S.\ James Gates,
Jr.,\footnote{sylvester${}_-$gates@brown.edu}$^{d}$ and Konstantinos
Koutrolikos\footnote{kkoutrolikos@physics.muni.cz}$^{e}$ }}
\\*[8mm]
\emph{
\centering
$^a$Department of Theoretical Physics,Tomsk State Pedagogical University,\\
Kievskaya, Tomsk 634041, Russia
\\[6pt]
$^b$National Research Tomsk State University,\\
Lenina, Tomsk 634050, Russia
\\[6pt]
$^c$Departamento de F\'isica, ICE, Universidade Federal de Juiz de Fora,\\
Campus Universit\'ario-Juiz de Fora, 36036-900, MG, Brazil
\\[6pt]
$^{d}$Department of Physics, Brown University,
\\[1pt]
Box 1843, 182 Hope Street, Barus \& Holley 545,
Providence, RI 02912, USA
\\[6pt]
$^e$ Institute for Theoretical Physics and Astrophysics, Masaryk
University,
\\[1pt]
Kotlarska, 611 37 Brno, Czech Republic
}
 $$~~$$
  $$~~$$
 \\*[-8mm]
{ ABSTRACT}\\[4mm]
\parbox{142mm}{\parindent=2pc\indent\baselineskip=14pt plus1pt
We give an explicit superspace construction of higher spin conserved supercurrents built out
of $4D,\N=1$ massless supermultiplets of arbitrary spin. These supercurrents are
gauge invariant and generate a large class of cubic interactions between a
massless supermultiplet with superspin $Y_1=s_1+1/2$ and two massless
supermultiplets of arbitrary superspin $Y_2$. These interactions are possible
only for $s_1\geq 2Y_2$. At the equality, the supercurrent acquires its
simplest form and defines the supersymmetric, higher spin
extension of the linearized Bel-Robinson tensor. 
}
\end{center}
$$~~$$
\vfill
\vfill
\clearpage
%
%%%%%%%%%%%%%%%%%%%%%%%%%%%%%%%%%%%%%%%%%%%%%%
\section{Introduction}
\label{intro}
%%%%%%%%%%%%%%%%%%%%%%%%%%%%%%%%%%%%%%%%%%%%%%
In previous works \cite{Superspace1,Superspace2,Superspace3,Superspace4,
Superspace5,Superspace6} various conserved, higher spin, supercurrent multiplets
of supersymmetric matter theories have been constructed. These were used
to generate first order interactions among matter theories and higher spin
supermultiplets with the lowest possible number of derivatives.
In this note, we go beyond matter theories and consider first order
interactions among higher spin supermultiplets.

Finding interactions involving higher spins is non-trivial as one has to 
satisfy the set of consistency conditions imposed by the gauge invariance of 
the free higher spin action. For non-supersymmetric theories the search for
higher spin interactions is extensive and in many cases interaction terms 
were successfully constructed for flat spacetime at first order in
coupling constant $g$ by using a variety of techniques, such as
light-cone approach \cite{lc1,lc2,lc3,lc4,lc5,lc6,lc7,lc8,lc9,lc10,lc11,lc12},
Noether's procedure \cite{Nm1,Nm2,Nm3,Nm4,Nm5} \footnote{Some of these
results were later generalized in \cite{Gen0,Gen1,Gen2,Gen3}. For a review see
\cite{review1}.} 
and
BRST \cite{BRST1,BRST2,BRST3,BRST4,BRST5,BRST6,BRST7,BRST8}.
Interestingly,
most of these results have been 
obtained by analysing tree level amplitudes of (super)strings
\cite{String1,String2,String3}, thus enhancing the connection
between string theory and higher spin fields. For (A)dS backgrounds
similar results have been obtained \cite{AdS1,AdS2,AdS3,AdS4,AdS5}
which eventually led to the fully interacting equations of motion
for higher spin fields \cite{fihs} (see \cite{rfihs} for a review).

Most of these constructions, focus on interaction vertices with the lowest
possible number of derivatives, which corresponds to a \emph{minimal coupling}
scenario. However we can have \emph{non-minimal} type of interactions which
in most cases will lead to lagrangians with higher derivatives. This has been
done in \cite{Nm3,Gen0} using gauge invariant field strengths. An interesting,
distinctive aspect of these interactions is their uniqueness up to trivial
redefinitions.

In this paper, we generalize these results to supersymmetric theories using the
manifest $4D,~\N=1$ standard superspace formulation \footnote{We use the
conventions of \emph{Superspace} \cite{GGRS}.}. 
Specifically, we give the explicit form of gauge invariant, higher spin, 
conserved supercurrents for all massless, $4D$ Minkowski, $\N=1$ higher spin 
supermultiplets. A particular example for the supergravity supercurrents
of low spins $j\leq1$ was given in \cite{BK}. Our results extends it in two
directions : (\emph{i}) we consider higher spin supermultiplets and
(\emph{ii}) we construct higher spin supercurrents.

Our construction is focused on the description of cubic interactions 
$Y_1-Y_2-Y_2$  between a
massless higher spin supermultiplet with arbitrary half-integer superspin 
$Y_1=s_1+1/2$ (which describes the propagation of massless spins $j=s_1+1$
and $j=s_1+1/2$) and two massless, higher spin supermultiplets with arbitrary
superspin
$Y_2$. We also assume that this cubic interaction can be written in the form
\emph{gauge suprefield times supercurrent}. The supercurrent must be quadratic
in the $Y_2$ supermultiplet and the gauge transformation of the $Y_1$ 
supermultiplet will impose on it an on-shell conservation equation. 
Additionally, we demand the supercurrent to be gauge invariant with respect to
the gauge transformation of the $Y_2$ supermultiplet. The $Y_2$-gauge 
invariance fixes the supercurrent to be quadratic in the derivatives of the 
$Y_2$ superfield strength and the conservation equation gives a unique solution
of this type. Furthermore, we derive a restriction on the allowed values of
superspin, $s_1\geq 2Y_2$, which provides a complete classification for this class of
interactions. This
is in agreement with the known constraints for the spin values
of higher spin interactions of non-supersymmetric theories \cite{Nm3,lc10}
as well as the Weinberg-Witten theorem \cite{WW}.
Right at the boundary, when $s_1=2Y_2$, the structure of the supercurrent
simplifies drastically. Its dependence on the superfield strength becomes
algebraic and defines the supersymmetric, higher spin extension of the 
Bel-Robinson\footnote{The Bel-Robinson tensor is known in the context
of General Relativity and is a spin 4, gauge
invariant and conserved tensor constructed out massless spin 2 fields. 
It is the generalization of the energy-momentum tensor.}
tensor \cite{MTW} (and reference therein).

The organization of the paper is as follows. In section 2, we review the
description of free, massless, $4D,~\N=1$ higher spin supermultiplets and their
corresponding superfield strengths. In section 3, we construct the gauge 
invariant, higher spin, conserved supercurrent which generates the cubic
interactions. In section 4, we include a component discussion for the
higher spin currents that can be extracted from the higher spin supercurrent
superfield. The last section summarizes our results.
%%%%%%%%%%%%%%%%%%%%%%%%%%%%%%%%%%%%%%%%%%%%%%%%%%%%%%%%%%%%%%%%%%%%%%%%%%%
\section{Free higher spin supermultiplets and their superfield strengths}
\label{sec2}
%%%%%%%%%%%%%%%%%%%%%%%%%%%%%%%%%%%%%%%%%%%%%%%%%%%%%%%%%%%%%%%%%%%%%%%%%%%
The massless, higher spin irreducible representations of the $4D,~\N=1$, 
super-Poincar\'{e}
were first described in \cite{hss1} using a component Lagrangian invariant under on-shell
supersymmetry.
%\footnote{It is instructive to mention that a classification of the
%irreducible representations of the $4D, \mathcal{N}=1$ super-Poincar\'{e}
%algebra was presented in \cite{ref1}.}.
Later a superfield
formulation was introduced in \cite{hss2,hss3,hss4} (see also
\cite{hss5,hss6,hss7}). A synopsis of the description of higher
spin supermultiplets is the following:
\begin{enumerate}
\item The integer superspin $Y=s$ ($s\geq1$) supermultiplets $(s+1/2 ,
s)$\footnote{On-shell they
describe the propagation of degrees of freedom with helicity $\pm (s+1/2)$ and
$\pm s$}
are described by a pair of superfields
$\Psi_{\a(s)\ad(s-1)}$\footnote{The notation $\a(k)$ is a shorthand for k
undotted symmetric indices $\a_1\a_2\dots\a_k$.
The same notation is used for the dotted indices }
 and $V_{\a(s-1)\ad(s-1)}$ (is real) with the following zero
order gauge transformations
\bea{l}\n\label{hstr1}
\d_0\Psi_{\a(s)\ad(s-1)}=-\D^2L_{\a(s)\ad(s-1)}+\tfrac{1}{(s-1)!}\Dd_{(\ad_{
s-1}}\Lambda_{\a(s)\ad(s-2))}~,\sn\\  \d_0 V_{\a(s-1)\ad(s-1)}=\D^{\a_s}L_{\a(
s)\ad(s-1)}+\Dd^{\ad_s}\bar{L}_{\a(s-1)\ad(s)}\sn~.
\eea
Off-shell, this supermultiplet carries $8s^2+8s+4$ bosonic and equal number of
fermionic degrees for freedom\footnote{A detailed counting of the off-shell
degrees of freedom can be found in \cite{ref2,hss5,hss6}.}.
%%%%
\item The half-integer superspin $Y=s+1/2$ supermultiplets $(s+1 , s+1/2)$
have two descriptions. The first is called the transverse formulation ($s\geq1$)
and it uses the pair of superfields $H_{\a(s)
\ad(s)}$ (is real) and $\chi_{\a(s)\ad(s-1)}$ with the following zero order gauge
transformations
\bea{l}\n\label{hstr2}
\d_0
H_{\a(s)\ad(s)}=\tfrac{1}{s!}\D_{(\a_s}\bar{L}_{\a(s-1))\ad(s)}-\tfrac{1}{s!}
\Dd_{(\ad_s}L_{\a(s)\ad(s-1))}\sn\label{hstr2H}~,\\
\d_0\chi_{\a(s)\ad(s-1)}=\Dd^2L_{\a(s)\ad(s-1)}+\D^{\a_{s+1}}\Lambda_{\a(
s+1)\ad(s-1)}~.\sn
\eea
This supermultiplet, off-shell describes $8s^2+8s+4$ bosonic and equal fermionic
degrees of freedom.
The second formulation is called the longitudinal ($s\geq2$) and it includes the
superfields $H_{\a(s)\ad(s)}$ (is real) and $\chi_{\a(s-1)\ad(
s-2)}$ with the following zero order gauge transformations
\bea{l}\n\label{hstr3}
\d_0
H_{\a(s)\ad(s)}=\tfrac{1}{s!}\D_{(\a_s}\bar{L}_{\a(s-1))\ad(s)}-\tfrac{1}{s!}
\Dd_{(\ad_s}L_{\a(s)\ad(s-1))}\sn~,\vspace{1ex}\\
\d_0\chi_{\a(s-1)\ad(s-2)}=\Dd^{\ad_{s-1}}\D^{\a_s}L_{\a(s)\ad(s-1)}+\tfrac{
s-1}{s}\D^{\a_s}\Dd^{\ad_{s-1}}L_{\a(s)\ad(s-1)}\sn\\
\hspace{17ex}
+\tfrac{1}{(s-2)!}\Dd_{(\ad_{s-2}}J_{\a(s-1)\ad(s-3))}~.
\eea
This supermultiplet carries $8s^2+4$ off-shell bosonic and equal number of
fermionic degrees of freedom. However one can show that it is dual to the
first, transverse, formulation\cite{hss2,Superspace6}.
\end{enumerate}

The physical and propagating degrees of freedom of the
two massless integer and half-integer superspin theories above are described by
corresponding superfield strengths $\W_{\a(2s)}$ and $\W_{\a(2s+1)}$
respectively. They are defined by:
\bea{ll}
Y=s+1/2 : ~&~\W_{\a(2s+1)}\sim\Dd^2\D_{(\a_{2s+1}}\pa_{\a_{2s}}{}^{\ad_s}
\pa_{\a_{2s-1}}{}^{\ad_{s-1}}\dots\pa_{\a_{s+1}}{}^{\ad_{1}}H_{\a(s))\ad(s)}\n
\vspace{2ex}\\
Y=s : &~\W_{\a(2s)}\sim\Dd^2\D_{(\a_{2s}}\pa_{\a_{2s-1}}{}^{\ad_{s-1}}
\pa_{\a_{2s-2}}{}^{\ad_{s-2}}\dots\pa_{\a_{s+1}}{}^{\ad_{1}}\Psi_{\a(s))\ad(s-1)}
\n
\eea
These are superfields which respect the gauge symmetries mentioned above and
at the component level they include the bosonic and fermionic higher spin field
strengths. Moreover, they have two interesting characteristics. Firstly, they
are both chiral superfields
\bea{l}
\Dd_{\bd}\W_{\a(2s+1)}=0~~~,~~~~\Dd_{\bd}\W_{\a(2s)}=0\n
\eea
Secondly, they have a special index structure. Specifically, they have indices
of only one type (the undotted ones) and in both cases the number of indices
is $2Y$. Furthermore, on-shell they satisfy the following equations of motions:
\bea{l}
\D^{\b}\W_{\b\a(2s)}=0~~~,~~~\D^{\b}\W_{\b\a(2s-1)}=0~.\n
\eea  
%%%%%%%%%%%%%%%%%%%%%%%%%%%%%%%%%%%%%%%%%%%%%%%%%%%%%%%%%%%%%%%
\section{Gauge invariant, conserved, higher spin supercurrent}
\label{sec3}
%%%%%%%%%%%%%%%%%%%%%%%%%%%%%%%%%%%%%%%%%%%%%%%%%%%%%%%%%%%%%%%
Now let us consider the cubic interaction $Y_1-Y_2-Y_2$ between a massless 
superspin $Y_1$ supermultiplet and two massless superspin $Y_2$ 
supermultiplets. We select $Y_1$ to be half-integer $Y_1=s_1+1/2$ for arbitrary
positive integer $s_1$ and $Y_2$ can be either integer, $Y_2=s_2$, or half-integer, 
$Y_2=s_2+1/2$, for arbitrary positive integer $s_2$. Moreover, we assume that
the first order interaction vertices can be written in a lagrangian form 
which is generated by a higher spin supercurrent $\J$ and a higher spin 
supertrace $\T$ as follows\footnote{Due to the duality between the transverse
and longitudinal formulations of the half-integer superspin, we only have to
consider one of them. Our choice is the transverse description.}
:
\bea{l}
S_{I}=g\int d^8z \left\{\vphantom{\frac12}
H^{\a(s_1)\ad(s_1)}\J_{\a(s_1)\ad(s_1)}+
\chi^{\a(s_1)\ad(s_1-1)}\T_{\a(s_1)\ad(s_1-1)}\n
\right\}~.
\eea
%In this case, because the $Y_2$ supermultiplets are not interacting the
%second term can be ignored because there are improvement terms that can
%eliminate the supertrace ($\T_{\a(s_1)\ad(s_1-1)}=0$). 
In this case, the supertrace $\T_{\a(s_1)\ad(s_1-1)}$ can be 
ignored because one can find improvement terms to eliminate it
($\T_{\a(s_1)\ad(s_1-1)}=0$). This is
related to the fact that a gauge invariant description of supermultiplet
$Y_2$ via the superfield strength has conformal symmetry. Hence the only
non-trivial object to focus in the supercurrent $\J_{\a(s_1)\ad(s_1)}$.

It is straightforward to show that, due to the gauge transformation
\eqref{hstr2H} the higher spin supercurrent must satisfy on-shell (up to terms
that depend on the equation of motion) the following conservation equation
\bea{l}
\Dd^{\ad_{s_1}}\J_{\a(s_1)\ad(s_1)}=0~.\n\label{ce}
\eea
Also due to the reality of $H_{\a(s_1)\ad(s_1)}$, $\J_{\a(s_1)\ad(s_1)}$ must
be real:
\bea{l}
\J_{\a(s_1)\ad(s_1)}=\mathcal{\bar{J}}_{\a(s_1)\ad(s_1)}~.\n\label{real}
\eea
On top of these requirements, we demand that the higher spin supercurrent is
gauge invariant with respect the gauge transformation of the $Y_2$ 
supermultiplets. This can be achieved if the supercurrent explicitly depends
on the superfield strength $\W_{\a(2Y_2)}$ of the $Y_2$ supermultiplet. Furthermore, due to the
cubic nature of the interaction the supercurrent must be quadratic in the
$Y_2$ information. However because of the special index structure of 
$\W_{\a(2Y_2)}$, we conclude that the supercurrent must explicitly depend on
both $\W_{\a(2Y_2)}$ and $\mathcal{\bar{W}}_{\ad(2Y_2)}$. The most general
ansatz one can write is:
\bea{l}
\J_{\a(s_1)\ad(s_1)}=\sum_{p=0}^{s_1-2Y_2}\a_{p}~\pa^{(p)}\W_{\a(2Y_2)}~
\pa^{(s_1-2Y_2-p)}\mathcal{\bar{W}}_{\ad(2Y_2)}\n\label{Jansatz}\\
\hspace{10ex}+\sum_{p=0}^{s_1-2Y_2-1}\b_{p}~\pa^{(p)}\D\W_{\a(2Y_2)}~
\pa^{(s_1-2Y_2-1-p)}\Dd\mathcal{\bar{W}}_{\ad(2Y_2)}
\eea
for some coefficients $\a_p$ and $\b_p$. We have intentionally not explicitly 
written
the external indices of the spacetime or spinorial derivatives in order to 
simplify the expression. Also one must not forget the independent symmetrization
of all the dotted and undotted indices of the right hand side together with
appropriate symmetrization factors that we also omit. The symbol 
$\pa^{(k)}$ is a replacement for a string of $k$ spacetime derivatives.

An immediate consequence of all the above is that there is a constraint
in the $Y_1$ and $Y_2$ values. A supercurrent of this type can exist only if:
\bea{l}
s_1\geq 2Y_2~~\Rightarrow~~Y_1\geq 2Y_2+1/2~~.\n\label{const}
\eea
In other words, if $Y_2>s_1/2$ there is not enough room for derivatives in order
to have a gauge invariant supercurrent. Hence the conclusion is that if a
supersymmetric theory allows the construction of a gauge invariant, conserved
(as in \eqref{ce}), real supercurrent of rank $s_1$ then its spectrum can 
include massless supermultiplets with superspin $Y>s_1/2$. This is
a supersymmetric, higher spin extension of the Weinberg-Witten
theorem \cite{WW}. Also at the component level the constrain \eqref{const}
is consistent with the restrictions found in \cite{Nm3,lc10}.

Now we have to actually check, whether a solution of (\ref{ce}) and (\ref{real})
of the form \eqref{Jansatz} exist. The conservation equation \eqref{ce} relates
the $\b_{p}$ and $\a_p$ coefficients:
\bea{l}
\b_{p}=-i~(-1)^{2Y_2}~\a_{p+1}~\frac{p+1}{s_1-p}~~,~~p=0,1,\dots ,s_1-2Y_2-1\n
\eea
whereas the reality of the supercurrent gives the following constraints:
\bea{l}
\a_{p}=\a^{*}_{s_1-2Y_2-p}~~,~~p=0,1,\dots ,s_1-2Y_2~,\n\vspace{1ex}\\
\b_{p}=\b^{*}_{s_1-2Y_2-1-p}~~,~~p=0,1,\dots ,s_1-2Y_2-1~.\n
\eea
The above three constraints fix uniquely the coefficients $\a_{p}$ and $\b_{p}$
\bea{l}
\a_{p}=c~(i)^{s_1-2Y_2}~(-1)^{p}~
\frac{\binom{s_1-2Y_2}{p}\binom{s}{p}}{\binom{2Y_2+p}{2Y_2}}~~,~~p=0,1,\dots ,
s_1-2Y_2~~,\n\vspace{2ex}\\
%%%%%%%%%%
\b_{p}=c~(i)^{s_1-2Y_2+1}~(-1)^{p+2Y_2}~
\frac{\binom{s_1-2Y_2}{p}\binom{s}{p}}{\binom{2Y_2+p}{2Y_2}}
\frac{s_1-2Y_2-p}{2Y_2+1+p}~~,~~p=0,1,\dots ,s_1-2Y_2-1\n
\eea
up to an overall real constant $c$. Hence not only is there a solution but it
is unique. The higher spin supercurrent we find is
\bea{l}
\J_{\a(s_1)\ad(s_1)}=c(i)^{s_1-2Y_2}\sum_{p=0}^{s_1-2Y_2}
(-1)^{p}\frac{\binom{s_1-2Y_2}{p}\binom{s}{p}}{\binom{2Y_2+p}{2Y_2}}
\left\{\vphantom{\frac12}
\pa^{(p)}\W_{\a(2Y_2)}~
\pa^{(s_1-2Y_2-p)}\mathcal{\bar{W}}_{\ad(2Y_2)}\right.\n\label{J}\\
\hspace{42ex}\left.\vphantom{\frac12}
+i(-1)^{2Y_2}~\frac{s_1-2Y_2-p}{2Y_2+1+p}
~\pa^{(p)}\D\W_{\a(2Y_2)}~
\pa^{(s_1-2Y_2-1-p)}\Dd\mathcal{\bar{W}}_{\ad(2Y_2)}
\right\}
\eea

There are two interesting cases to look at. The first one is at the boundary of
\eqref{const}, when $s_1=2Y_2$. In this case the supercurrent \eqref{J} is
simplified and takes the following form:
\bea{l}
\J_{\a(s_1)\ad(s_1)}=c~ \W_{\a(2Y_2)}\mathcal{\bar{W}}_{\ad(2Y_2)}~.\n\label{BR}
\eea
This supercurrent is the supersymmetric and higher spin extension of the
Bel-Robinson tensor. That means that for $Y_2=3/2$ (linearized supergravity)
the supercurrent $\J_{\a\b\g\ad\bd\gd}$ as defined by \eqref{BR} in its 
components includes the Bel-Robinson tensor of General Relativity
(see \cite{MTW} and references their in). For higher values of $Y_2$ we get
higher spin generalizations of the Bel-Robinson tensor. A supercurrent of this
type has been constructed in \cite{ref3}.

The second interesting observation is in the limit $Y_2=0$. This case technically
is not allowed because it is not a gauge theory but a matter theory, hence
there is no zero order transformation and $\W$ is not a superfield strength.
However, it is the chiral superfield of the free chiral theory and the 
suppercurrent \eqref{J} gives the cubic interactions of the higher spin 
supermultiplet $Y_1$ with the chiral supermultiplet. Therefore it is allowed
to take this limit. In that case the expression \eqref{J} exactly matches to the
higher spin supercurrent of matter theories found in 
\cite{Superspace1,Superspace2}.
%%%%%%%%%%%%%%%%%%%%%%%%%%%%%%%%%%%%%%%%%%%%%%%%%%%%%%%%%%%%
\section{Component structure of the higher spin supercurrent}
\label{sec4}
%%%%%%%%%%%%%%%%%%%%%%%%%%%%%%%%%%%%%%%%%%%%%%%%%%%%%%%%%%%%
The components of the higher spin supercurrent \eqref{J} will include bosonic
and fermionic gauge invariant, conserved higher spin currents. It will be
useful to extract the expressions for these currents and compare with known 
results for non-supersymmetric theories. For the projection of the superfield
to its components, we will follow \cite{hss5,Superspace2}.
Due to the conservation equation \eqref{ce},  on-shell the higher spin
supercurrent $\J_{\a(s_1)\ad(s_1)}$ has only three independent components
\footnote{The labels $(0,0)$, $(1,0)$ and $(1,1)$ correspond to the position of
the components in the $\theta$ and $\thd$ expansion of the supercurrent.}
(2 bosonic and one fermionic)
\bea{l}
\J^{(0,0)}_{\a(s_1)\ad(s_1)}=\J_{\a(s_1)\ad(s_1)}
\Bigr|_{\substack{\th=0\\ \thd=0}}~,\hspace{2ex}
%%%%%%%%%%%%%%%%%%%%%%%%
\J^{(1,0)}_{\a(s_1+1)\ad(s_1)}=\tfrac{1}{(s_1+1)!}
\D_{(\a_{s_1+1}}\J_{\a(s_1))\ad(s_1)}\Bigr|_{\substack{\th=0\\ \thd=0}}~,\n\\
%%%%%%%%%%%%%%%%%%%%%%%%
\J^{(1,1)}_{\a(s_1+1)\ad(s_1+1)}=-\tfrac{1}{2(s_1+1)!(s_1+1)!}
[\D_{(\a_{s_1+1}},\Dd_{(\ad_{s_1+1}}]
\J_{\a(s_1))\ad(s_1))}\Bigr|_{\substack{\th=0\\ \thd=0}}~.
\eea
The rest of them either vanish or are derivatives of the components above.
It is straight forward to show, using \eqref{real} and \eqref{ce}, that these
components satisfy the traditional spacetime conservation equations
\bea{l}
\pa^{\a_{s_1}\ad_{s_1}}\J^{(0,0)}_{\a(s_1)\ad(s_1)}=0,~~
\pa^{\a_{s_1+1}\ad_{s_1}}\J^{(1,0)}_{\a(s_1+1)\ad(s_1)}=0,~~
\pa^{\a_{s_1+1}\ad_{s_1+1}}\J^{(1,1)}_{\a(s_1+1)\ad(s_1+1)}=0\n
\eea
hence they correspond to gauge invariant, conserved, higher spin currents.
Their explicit expressions are:
\bea{l}\n\label{Jcurrent}
\J^{(0,0)}_{\a(s_1)\ad(s_1)}\sim
(i)^{s_1-2Y_2}\sum_{p=0}^{s_1-2Y_2}
(-1)^{p}\frac{\binom{s_1-2Y_2}{p}\binom{s}{p}}{\binom{2Y_2+p}{2Y_2}}
\left\{\vphantom{\frac12}
\pa^{(p)}\W^{(0,0)}_{\a(2Y_2)}~
\pa^{(s_1-2Y_2-p)}\mathcal{\bar{W}}^{(0,0)}_{\ad(2Y_2)}\right.\sn\label{J00}\\
\hspace{39ex}\left.\vphantom{\frac12}
+i(-1)^{2Y_2}~\tfrac{s_1-2Y_2-p}{2Y_2+1+p}
~\pa^{(p)}\W^{(1,0)}_{\a(2Y_2+1)}~
\pa^{(s_1-2Y_2-1-p)}\mathcal{\bar{W}}^{(0,1)}_{\ad(2Y_2+1)}
\right\}~,\vspace{1ex}\\
%%%%%%%%%%%%%%%%%%%%%%%%%%%%%%%%%%
%%%%%%%%%%%%%%%%%%%%%%%%%%%%%%%%%%
\J^{(1,0)}_{\a(s_1+1)\ad(s_1)}\sim
(i)^{s_1-2Y_2}\sum_{p=0}^{s_1-2Y_2}
(-1)^{p}\frac{\binom{s_1-2Y_2}{p}\binom{s}{p}}{\binom{2Y_2+p}{2Y_2}}
\tfrac{s_1+1}{2Y_2+1+p}~
\pa^{(p)}\W^{(1,0)}_{\a(2Y_2+1)}~~
\pa^{(s_1-2Y_2-p)}\mathcal{\bar{W}}^{(0,0)}_{\ad(2Y_2)}\sn\label{J10}
~,\vspace{1.5ex}\\
%%%%%%%%%%%%%%%%%%%%%%%%%%%%%%%%%%
%%%%%%%%%%%%%%%%%%%%%%%%%%%%%%%%%%
\J^{(1,1)}_{\a(s_1+1)\ad(s_1+1)}\sim
(i)^{s_1-2Y_2}\sum_{p=0}^{s_1-2Y_2}
(-1)^{p}\frac{\binom{s_1-2Y_2}{p}\binom{s}{p}}{\binom{2Y_2+p}{2Y_2}}
\left\{\vphantom{\frac12}
i\pa^{(p)}\W^{(0,0)}_{\a(2Y_2)}~
\pa^{(s_1+1-2Y_2-p)}\mathcal{\bar{W}}^{(0,0)}_{\ad(2Y_2)}\right.\sn\label{J11}\\
\hspace{52ex}-i~\tfrac{2s_1-2Y_2+1-p}{2Y_2+1+p}~
\pa^{(p+1)}\W^{(0,0)}_{\a(2Y_2)}~
\pa^{(s_1-2Y_2-p)}\mathcal{\bar{W}}^{(0,0)}_{\ad(2Y_2)}\vspace{0.6ex}\\
\hspace{43ex}+(-1)^{2Y_2}~\tfrac{s_1+2Y_2+2+p}{2Y_2+1+p}~
\pa^{(p)}\W^{(1,0)}_{\a(2Y_2+1)}~
\pa^{(s_1-2Y_2-p)}\mathcal{\bar{W}}^{(0,1)}_{\ad(2Y_2+1)}\vspace{0.6ex}\\
\hspace{41ex}\left.\vphantom{\frac12}
-(-1)^{2Y_2}~\tfrac{s_1-2Y_2-p}{2Y_2+1+p}
~\pa^{(p+1)}\W^{(1,0)}_{\a(2Y_2+1)}~
\pa^{(s_1-2Y_2-1-p)}\mathcal{\bar{W}}^{(0,1)}_{\ad(2Y_2+1)}
\right\}~.
\eea
The components $\W^{(0,0)}_{\a(2Y_2)}$ and $\W^{(1,0)}_{\a(2Y_2+1)}$
are the only non-trivial components of the superfield strength $\W_{\a(2Y_2)}$ 
\bea{l}
\W^{(0,0)}_{\a(2Y_2)}=\W_{\a(2Y_2)}
\Bigr|_{\substack{\th=0\\ \thd=0}}~~,~~
%%%%%%%%%%%%%%%%%%%%%%%%
\W^{(1,0)}_{\a(2Y_2+1)}=\tfrac{1}{(2Y_2+1)!}
\D_{(\a_{2Y_2+1}}\W_{\a(2Y_2))}\Bigr|_{\substack{\th=0\\ \thd=0}}\n
\eea
and are the higher spin field strengths for spins $j=Y_2$ and $j=Y_2+1/2$
respectively. These \eqref{Jcurrent} higher spin currents are consistent
with the results of \cite{Nm3,Gen0}.
%%%%%%%%%%%%%%%%%%%%%%%%%%%%%%%%%%%%%%%%%%%%%%%%%
\section{Discussion}
\label{disc}
%%%%%%%%%%%%%%%%%%%%%%%%%%%%%%%%%%%%%%%%%%%%%%%%%
We consider cubic interactions $Y_1-Y_2-Y_2$ among massless 
supermultiplets of superspin $Y_1$ and $Y_2$, where $Y_1$ is half integer
$Y_1=s_1+1/2$ for arbitrary positive integer $s_1$ and $Y_2$ has an arbitrary
non-negative integer or half-integer value. Specifically, we consider the class of these interactions
that are generated by a higher spin supercurrent $\J_{\a(s_1)\ad(s_1)}$ which
respects the gauge symmetry of the $Y_2$ supermultiplet. We find that:
\begin{enumerate}
\item[\emph{i}.] There is a unique gauge invariant, conserved,
real, higher spin supercurrent $\J_{\a(s_1)\ad(s_1)}$ \eqref{J}.
\item[\emph{ii}.] The existence of this supercurrent puts constraints 
\eqref{const} on the allowed $Y_2$ superspin values [$s_1\geq 2Y_2$]. This is a 
supersymmetric and higher spin generalization of the Weinberg-Witten theorem 
and also is consistent with various spin restrictions coming from the 
consideration of non-supersymmetric, higher spin, cubic interactions.
\item[\emph{iii}.] For the special case of $s_1=2Y_2$ the higher spin
supercurrent simplifies a lot \eqref{BR} and agrees with the results of \cite{ref3}.
This supercurrent provides a supersymmetric, higher
spin extension of the linearized Bel-Robinson tensor known in the context of General
Relativity.
\item[\emph{iv}.] Another special limit is $Y_2=0$ where we recover previously
known results for the higher spin supercurrent of the free, massless chiral theory.
\item[\emph{v}.] The component structure includes two bosonic and one fermionic
gauge invariant, conserved, higher spin currents \eqref{Jcurrent} which depend
on the derivatives of higher spin field strengths for spins $j=Y_2$ and 
$j=Y_2+1/2$.
\end{enumerate}
The list of the supercurrents we find refer only to non-minimal
couplings (higher number of derivatives) and that is
due to the explicit dependence of the supercurrent on the superfield strengths.
Therefore, in spite of this list being infinite (for arbitrary values of $Y_1$ and
$Y_2$ that respect \eqref{const}), it is incomplete.
This is simply understood
from the fact that ordinary conserved currents such as
the energy momentum tensor and their higher spin extensions \cite{nGI-hsc}
are not included in the
class of gauge invariant currents.
It would be interesting to search for 
non gauge invariant higher spin supercurrents
that generate minimal coupling (least
number of derivatives) among higher spins and give gauge invariant
conserved charges.

%%%%%%%%%%%%%%%%%%%%%%%%%%%%%%%%%%%%%%%%%%%%%%%
%%%%%%%%%% Acknowledgement %%%%%%%%%%%%%%%%%%%%
{\bf Acknowledgements}\\[.1in] \indent
The research of I.\ L.\ B.\ was supported in parts by
Russian Ministry of Education and Science, project No. 3.1386.2017.
He is also grateful to RFBR grant, project No. 18-02-00153 for
partial support. The research of S.\ J.\ G.\ is supported by the 
endowment of the Ford Foundation Professorship of Physics at 
Brown University. The work of K.\ K.\ was supported by the grant
P201/12/G028 of the Grant agency of Czech Republic.
%%%%%%%%%%%%%%%%%%%%%%%%%%%%%%%%%%%%
%%%%%%%%% Bibliography %%%%%%%%%%%%%%%%%%%%

\end{document}